\documentclass[prb,preprint]{revtex4-1} 

\usepackage{amsmath}  % needed for \tfrac, \bmatrix, etc.
\usepackage{amsfonts} % needed for bold Greek, Fraktur, and blackboard bold
\usepackage{graphicx} % needed for figures

\newcommand{\eq}[1]{\begin{gather} #1 \end{gather}}
\newcommand{\eqa}[1]{\begin{align} #1 \end{align}}
\newcommand{\vc}[1]{{ \boldsymbol #1 }}

\newcommand{\fr}[1]{\ref{fig:#1}}
\newcommand{\er}[1]{(\ref{eq:#1})}

\newcommand{\sr}[1]{\ref{sec:#1}}

\begin{document}

% Be sure to use the \title, \author, \affiliation, and \abstract macros
% to format your title page.  Don't use lower-level macros to  manually
% adjust the fonts and centering.

\title{Observation of the Talbot effect with water waves}
% In a long title you can use \\ to force a line break at a certain location.

\author{Alexandra Bakman}
\affiliation{Department of Physics,Technion- Israel Institute of Technology, Haifa, Israel 3200003}

\author{Shmuel Fishman}
\affiliation{Department of Physics,Technion- Israel Institute of Technology, Haifa, Israel 3200003}

\author{Mathias Fink}
\affiliation{Institut Langevin, ESPCI Paris, PSL Research University, CNRS UMR 7587, 1 rue Jussieu, 75238 Paris, France}

\author{Emmanuel Fort}
\affiliation{Institut Langevin, ESPCI Paris, PSL Research University, CNRS UMR 7587, 1 rue Jussieu, 75238 Paris, France}
%\email{emmanuel.fort@espci.fr}

\author{Sander Wildeman}
\affiliation{Institut Langevin, ESPCI Paris, PSL Research University, CNRS UMR 7587, 1 rue Jussieu, 75238 Paris, France}
\affiliation{PMMH, ESPCI Paris, CNRS, PSL Research University, 1 rue Jussieu, F-75005 Paris, France}
\email{swildeman@gmail.com}

\begin{abstract}
When light is incident upon a diffraction grating, images of the grating appear at periodic intervals behind the grating. This phenomenon and the associated self-imaging distance were named after Talbot who first observed them in the nineteenth century. A century later, this effect held new surprises with the discovery of sub-images at regular fractional distances of the Talbot length. In this paper, we show that water waves enable one to observe the Talbot effect in a classroom experiment. Quantitative measurements, of for example the Talbot distances, can be performed with an easy to use digital Schlieren method. 
\end{abstract}

\maketitle % title page is now complete

\section{Introduction}

The Talbot effect is a fascinating effect observed in the simple configuration of a grating illuminated by coherent light. A series of self-images are produced at regular distances behind the grating. This phenomenon was first observed by Talbot in 1836.\cite{Talbot1836} Lord Rayleigh gave a first explanation of the Talbot effect as a result of the interference between the diffraction orders.\cite{Rayleigh1881} He showed that the distances of the images from the grating were a multiple of the so called Talbot distance $l_T = d^2/\lambda$, where $d$ is the period of the grating and $\lambda$ the wavelength of the light. 

This seemingly simple phenomena still holds new surprises.  The subject was revived after more than 80 years with the discovery of fractional Talbot images at rational multiples of $l_T$, i.e. for distances $(p/q)l_T$ with $p$ and $q$ being coprime integers.\cite{Hiedemann1959,Winthrop1965} More recently, Berry and Klein gave a general framework and demonstrated the presence of a fractal behavior for intensity profiles at irrational distances from the grating and along slices not parallel to the grating. \cite{Berry1996}

The Talbot effect is generic of coherent wave interference and thus not confined to light waves. It has been demonstrated with matter waves of BEC condensate,\cite{Deng1999,Gammal2004} metastable helium atoms, \cite{Nowak1997} photonic crystals\cite{Zhang2010} and rogue waves.\cite{Zhang2014,Zhang2015} Its quantum analog is the space-time quantum carpet woven by the probability density of a particle in a box.\cite{Marzoli1998,Berry2001} 

In this article, we show that the Talbot effect and its fractional version can be observed with the naked eye in a classroom experiment using water waves. 

Water waves offer several unique advantages for introducing wave phenomena to students. Because of their slow velocity and macroscopic wavelengths, their propagation can be observed with standard cameras and controlled dynamically with characteristic times that are much smaller than the propagation time. In addition, their dispersion relation can be engineered using bathymetry and it is possible to design specific geometries and boundaries with a sub-wavelength resolution. Finally, water waves couple strongly to floating objects and exert a large radiation pressure on the cavity walls. These features  have been used to introduce original phenomena like a classical analog of the wave-particle duality,\cite{Couder2010} time-reversal mirrors \cite{Bacot2016} or cavity self-adaptation under wave pressure.\cite{Pucci2011} In relation to the Talbot effect, a recent experiment used water waves to show a self-imaging effect in the case of a parametrically excited fluid.\cite{Sungar2017} In this case, however, the wavelength selection driven by the excitation instability (Faraday instability) prevents the appearance of fractional images.
 
This paper is organized as follows: in the next section we first introduce a geometrical interpretation of the Talbot effect, followed by a recapitulation the standard wave theory (in which we highlight the differences with the geometrical picture). Section \sr{method} describes the experimental setup and the wave measurement technique. Section \sr{results} is devoted to the results and discussion and we conclude with some ideas of how the setup could be adapted for other wave physics demonstrations.

\section{Theory}

\subsection{Geometric interpretation}
\label{sec:geom}

When a coherent beam of light from a laser impinges on a grating, wave interference causes the beam to split up in multiple diffracted beams. The angles $\theta_n$ at which these secondary beams emerge is given by the familiar grating formula
\eq{
	\sin(\theta_n) = \frac{n\lambda}{d} \qquad (n \in 0, \pm 1, \pm 2, ...), \label{eq:grat}
}
where $\lambda$ is wavelength of the light and $d$ is the distance between the slits. This formula expresses the condition that the path difference between light rays from adjacent slits should be an integer number of wavelengths for constructive interference to occur (see Fig. \fr{theory}(a)). At distances large compared to the size of the grating (in the far field) these diffracted beams show up as distinct spots on the wall. However, as visualized in Fig. \fr{theory}(b), just behind the grating the different diffraction orders overlap, giving rise to an intricately woven interference pattern in which sharply focused images of the grating re-appear at regular intervals. This is the essence of the Talbot effect.

Figure \fr{theory}(b) was constructed by drawing rays according to the grating formula (Eq. \er{grat}) from eleven equispaced points representing the slits. Although this simple ray construction does not do complete justice to the wave nature of the phenomenon (as we will see below), it reproduces many of its basic geometric features. It suggests, for example, that the first re-focusing appears at a distance where the 1st-order diffraction ray ($n = \pm 1$), drawn from a given slit, intersects with the zeroth-order ray from its neighbor. This happens at a distance
\eq{
	l_T \approx \frac{d}{\theta_1} \approx \frac{d^2}{\lambda},
}
called the Talbot distance.\cite{Rayleigh1881} This formula holds in the small angle approximation, where $\sin(\theta) \approx \tan(\theta) \approx \theta$. In this approximation, higher order rays from subsequent neighbors all intersect in the same point, forming a sharp focus. For gratings that scatter a significant part of the incident beam into larger angles the pattern will be less well-defined. 

The ray construction also reveals some interesting fine-structure, such as the appearance of images at half the Talbot distance and other fractional distances $(p/q) l_T$. These images can be understood as the intersections of the $\pm n$th order ray from one slit with the $\mp (q-n)$th order ray from the $p$th nearest neighbor (where, for each distance, $n$ runs from $0$ to $q-1$). For example, at half the Talbot distance ($p = 1$, $q = 2$) two images per grating period are observed, at $(1/3)l_T$ and $(2/3)l_T$ three images per period, etc. The more diffraction orders are included (while still satisfying the small angle approximation) the more intricate the pattern will become.

If we compare the distance of the zeroth order ray to the first focus, $l_0 = l_T \approx d^2/\lambda$, to that of the higher order rays to the same focus, $l_n \approx \sqrt{(d^2/\lambda)^2 + (n d)^2} \approx d^2/\lambda +n^2 \lambda /2$, we see that the difference is a multiple of \emph{half} a wavelength, leading to destructive interference between adjacent neighbors. So instead of a maximum in intensity, as expected from the ray picture, we find a minimum if we include interference effects. In fact, one can show that constructive interferences occurs at positions vertically shifted by $d/2$, right in between the original foci. The first focus that reproduces an unshifted image of the grating occurs at \emph{twice} the Talbot distance, i.e. at $2 d^2/\lambda$. As we will see below, these findings are recovered in the more complete wave theory.

\subsection{Wave theory}

Following Rayleigh and later investigators [2,3,4], we now turn to a more rigorous wave description of the phenomenon.  We will take the $y$-axis to coincide with the grating, while the incoming wave propagates normal to this, in the $x$-direction. Just behind the grating, at $x = 0$, the wave field can be approximated as if it originated from an infinite periodic array of slits. This implies that here the wave amplitude can be expanded in a Fourier series in which the wave vectors $k_y$ along the grating can only take values which are integer multiples of $2\pi/d$.  Since the magnitude of the wave vector, $k =  \sqrt{k_x^2 + k_y^2} = 2\pi/\lambda$, is conserved in the experiment (because we operate at a fixed frequency $\omega(k)$) the condition on $k_y$ also determines $k_x$. Notice that the same discrete spectrum for $k_y$ can also be inferred directly from the grating formula (Eq. \er{grat}), since $k_y = k \sin(\theta) =  2\pi n/ d$. 

The wave field emerging from the grating can be explicitly written as
\eq{
	a(x,y,t) = e^{-i \omega(k) t}\sum_{n=-\infty}^{\infty} c_n \exp\left(i \frac{2 \pi n}{d} y + i \sqrt{\left(\frac{2\pi}{\lambda}\right)^2 - \left(\frac{2 \pi n}{d}\right)^2} x \right) \label{eq:series}
}
where $c_n$ are expansion coefficients that depend on the shape of the slits and the transmission process. Roughly the $c_n$ can be taken as the Fourier series coefficients that describe the transparency of the grating. For a series of slits of width $w$, we have for example $c_0 = w/d$ and $c_n =\sin(n\pi w/d)/(n \pi)$ for $n \neq 0$.

The expansion coefficients $c_n$ describing the grating quickly fall off as the argument of the sine function approaches $\pi$. This means that most of the power will be in the diffraction orders for which
\eq{
|n| \lesssim d/w. \label{eq:condw}
}
In addition, only $n$ for which the square root in Eq. \er{series} remains real, i.e.
\eq{
|n| < d/\lambda, \label{eq:condl}
}
will contribute to the Talbot pattern. Other terms decay to zero at a short distance (of the order of the wavelength) away from the grating. It follows that, to have enough diffraction orders to reproduce a sharp image of the grating at some distance from the grating, we should choose $\lambda \ll w$. This also ensures that most of the power is radiated at small angles $\theta < \lambda/w$, so that a small angle approximation (essential in the derivations that follow) is justified.

In Fig. \fr{theory}(c) we plotted the intensity pattern $|a|^2 = aa^*$ calculated using Eq. \er{series} for a grating with $w/d = 0.1$ and a wavelength such that $\lambda/w = 0.1$. For this choice of parameters, i.e. $\lambda \ll w$ and $w \ll d$, the picture that we obtain shares many of the features of the geometric interpretation discussed above. In particular, we observe sharp images of the grating at distances which are integer multiples of $l_T$. Also the repeated images at fractional distances are clearly visible.

To understand these features directly from Eq. \er{series}, it is essential (as it was for the geometric case) to work in the small angle (or paraxial) approximation. In this approximation $n\lambda \ll d$, so that we can expand the square root in Eq. \er{series} to first order in $(n \lambda/d)^2$. After some rearrangements, we obtain:
\eq{
	a(x,y,t) = e^{i\frac{2\pi}{\lambda}x-i \omega(k) t}\sum_{n=-\infty}^{\infty} c_n \exp\left(2\pi i \left[ \frac{n}{d} y - \frac{ \lambda n^2}{2 d^2} x\right] \right).
}
In this form, it is immediately clear that (up to an overall phase factor) the field amplitude (and thus its intensity envelope, $|a|^2 = aa^*$) is a periodic function in $x$, with a period of $2d^2/\lambda$, \emph{two} times the Talbot distance. 

Another interesting situation occurs at $x = l_T$. In this case each term in the series is multiplied by a factor $\exp(-i\pi n^2)$, which takes the value $1$ or $-1$ depending on whether $n^2$ is even or odd. Since $n^2$ and $n$ behave in the same way in this respect (that is, they have the same parity), we can just as well write the exponential factor as $\exp(-i\pi n)$. We then obtain
\eqa{
	a(l_T,y,t) &= e^{i\frac{2\pi}{\lambda}x-i \omega(k) t}\sum_{n=-\infty}^{\infty} c_n \exp\left(i \frac{2 \pi n}{d} (y - d/2) \right),
}
which is just the field at $x = 0$ shifted vertically by $d/2$. This matches with our earlier observation that \emph{at} the Talbot distance the grating pattern is reproduced, but with a shift. Here we see that, whatever the transmission function of our grating (represented by the $c_n$), a perfect image of it is produced at integer multiples of $l_T$ (in the paraxial approximation, that is). 

By similar mathematical operations one can show that additional, weakened copies of the grating transmission function show up at the fractional distances $(p/q) l_T$.\cite{Hiedemann1959} Since (as shown in the previous section) the vertical spacing between these fractional copies reduces with increasing $q$, those copies can (partially) overlap if the width of the slits is large enough. For a transmission function with sharp edges, the rugged intensity profile resulting from such a collection of overlapping and (coherently) interfering images can have interesting fractal properties.\cite{Berry1996} In our experiments, however, we restricted our attention to the more geometrical situation with $w \ll d$.

\section{Experimental method}
\label{sec:method}

Our experimental setup consists of a $370\times370\times20$ mm$^3$ plexiglas basin filled to a height of $15$ mm with plain tap water. 
As shown schematically in Fig. \fr{exp}(a), a horizontally oscillating paddle spanning the full width of the container acts as a source of plane waves that impinge onto a grating. The grating is placed parallel to wave fronts and consists of a thin plexiglas plate with 15 rectangular slots where the waves can pass through. The spacing between the slots is $d = 20$\,mm and each slot is $w = 6$\,mm wide. A $10$\,mm long ``beach'' of shallow water was created near the walls of the basin to somewhat dampen reflections. The wave field behind the grating was recorded with a digital video camera (Basler acA1300-200um), viewing the water surface from above. The camera frame rate was set to 100\,Hz, high enough to capture the wave frequencies in our experiment, $f_{exp} = 30$ -- $49$\,Hz, without sampling artifacts. 

To quantify the local amplitude of the waves, a checkerboard pattern with 1\,mm wide squares (printed on paper) was placed below the transparent basin. When a wave travels over the water surface, the checkerboard pattern (viewed through the water) appears distorted. For small slopes of the water waves, the virtual displacement $\vc u(x,y)$ seen at a given location ($x$,$y$) on the pattern is $\vc u \approx -H (1 - n_a/n_w) \nabla \eta$,\cite{Moisy2009} where $\eta(x,y)$ is the local amplitude of the water wave, $H$ is the distance between the pattern and the water surface and $n_a/n_w \approx 0.75$ is the refractive index contrast between air and water. So by measuring the distortions in the recorded checkerboard images, the slope of the water surface and (after integration) its amplitude can be recovered. 

The distortions were extracted digitally from the raw footage by using the Fast Checkerboard Demodulation (FCD) technique, which was recently developed for this purpose.\cite{Wildeman2018} In this method the checkerboard pattern is treated as a high-frequency (2D) carrier wave which is modulated by the optical distortions caused by the water ripples (akin to FM radio, where an acoustic signal modulates a radio frequency carrier wave). Basic (spatial) Fourier domain filtering can then be used to extract the signal. Matlab scripts that implement the FCD method are freely available online.\cite{Wildeman2018}

\section{Results and discussion}
\label{sec:results}

Figure \fr{exp}(b) shows an example of a wave field obtained behind the grating when the paddle was operated at a frequency of $f = 40$\,Hz. To improve the signal to noise ratio, we filtered the movie of raw wave fields (obtained from the FCD analysis) in the frequency domain, keeping only the wave motions occurring at the driving frequency of $40$\,Hz. For clarity of presentation, the image was cropped around the central slots. The same wave field is repeated outside this region, up till about 2 slots from the edges of the grating, where the finite size of the grating comes into play.

Upon careful inspection of the wave field in Fig. \fr{exp}(b) one can already observe some of the repeated patterns associated with the Talbot effect. In Fig. \fr{exp}(d) we computed the 2D spatial Fourier transform of the wave field. Here the different diffraction orders are clearly visible as peaks in the spectral intensity. Notice that for the second and especially the third order beams a small angle approximation (assumed in the theoretical analysis above) is not really justified. As we will discuss below, this leads to some deviations from the ideal Talbot pattern.

The Fourier peaks lay on a circle with radius $k = 2\pi/\lambda$, where $\lambda$ is the wavelength of the incoming wave. From Fig. \fr{exp}(b) we obtain $k \approx 1000$\,rad/m, corresponding to a wavelength of $6.3$\,mm. This is close to the theoretical expected value, $k = 920$\,rad/m ($\lambda = 6.9$\,mm), obtained from the dispersion relation of gravity-capillary waves, $\omega(k) = \sqrt{(g k + (\gamma/\rho)k^3)\tanh(k D)}$,\cite{Milne-Thomson1996} where $g = 9.8$\,m/s$^2$ is the gravitational acceleration, $D = 15$\,mm is the water depth, and $\rho = 1000$\,kg/m$^3$ and $\gamma = 72$\,mN/m are, respectively, the density and surface tension of water. The slightly lower wavelength observed experimentally is probably due to surfactants at the water-air interface, which can significantly lower the surface tension from its value for pure water. We found that taking a surface tension of $\gamma \approx 55$\,mN/m results in a good match for all experimental frequencies.  

Instead of looking at the wave \emph{amplitude}, a more clear picture can be obtained by calculating the envelope or \emph{intensity} of the wave field (as was done the theory). If a sufficiently long movie is recorded at a sufficiently high frame rate (or by stroboscopic means, at a frame rate slightly out of step with the paddle frequency) this envelope can be obtained by simply calculating the maximum amplitude (in time) at each pixel in the image. A slightly more sophisticated way, which requires less data to be recorded, is to split the real valued wave field into two complex valued fields ($a(x,y) e^{-2 \pi i f t}$ and $a^*(x,y) e^{2 \pi i f t}$) using a Fourier transform and then to calculate $|a|^2 = a a^*$. We used the latter method to calculate the intensity profile shown in Fig. \fr{exp}(d). To bring out the features far from the grating, we normalized the intensity profile by the square of measured viscous decay shown in Fig. \fr{exp}(c). This works well up to a distance of about $150$\,mm from the grating. After that, the amplitude of the waves becomes too small to be properly detected by our method. By using clean distilled water instead of tap water the damping effect could probably be somewhat reduced, although we expect that even with this change it will be difficult to observe the waves over distances of more than about three Talbot lengths from the grating. On the other hand, the decay of the waves allows one to use a relatively small water tank without having to worry about reflections from the walls.

Although in our experiment the paraxial approximation is not satisfied very well (in our frequency range $\lambda \sim w$ instead of $\lambda \ll w$), some important features of the ideal geometric Talbot effect (Fig.\fr{theory}(c)) are clearly visible in the intensity profile in Fig. \fr{exp}(e). One can observe the appearance of sharp foci that mimic the grating, and even some fractional images are visible.

In Fig. \fr{results}(a) we have plotted the horizontal position of the two strongest foci as a function of the wavelength of the impinging water waves. Here we see that, for the larger wavelengths, the position of the foci does not exactly match with the prediction of the paraxial approximation, $n l_T = nd^2/\lambda$. In addition, instead of a single sharp focus, we observe a horizontal splitting of the focal points (see Fig. \fr{results}(b)). The location of the peak intensity cycles between these sub-foci as the wavelength is decreased, which explains the sudden jumps in focal distance in Fig. \fr{results}(a). This splitting phenomenon (which is also visible in the calculated pattern in Fig. \fr{results}(c)) is related to the fact that for these cases $\lambda/w \sim 1$, instead of $\lambda/w \ll 1$ as assumed in the paraxial approximation. This makes that, according to equation \er{series}, the horizontal repetition of the first order pattern (resulting from interference between diffraction orders $n=0$ and $n=\pm 1$) does not commensurate with that of the second order pattern (from interference between $n=0$ and $n=\pm 2$), resulting in a kind of beating between these two patterns when the wavelength is varied. As the wavelength is decreased, more diffraction orders contribute to the pattern, resulting in a smaller distance between the split intensity peaks (and correspondingly smaller jumps in Fig. \fr{results}(a)). For the smallest wavelength in our experiments, $\lambda \approx 5.4$\,mm, (see Fig. \fr{results}(d-e)) the splitting almost disappears and the focal distance approaches that of the paraxial theory.

Finally, we find a very good match, both qualitatively and quantitatively, between the experimental intensity profiles (Fig. \fr{results}(b) and (d)) and the simulated profiles (Fig. \fr{results}(c) and (e)) calculated using Eq. \er{series} (without approximations).

\section{Conclusion and outlook}

This paper shows that water waves are particularly well suited to introduce the Talbot effect into the classroom. The effect is of great interest (both from an application and from an educational perspective) because it introduces the ability to image without a lens, which although non-intuitive at first sight belongs to the general class of coherent wave control devices, ranging from Fresnel zone plates to spatial phase modulators. 

The macroscopic wavelengths and the slow propagation speed of water waves enable one to observe the wave field in real time with the naked eye. By using a stroboscopic light (operating at the paddle frequency) wave fields qualitatively similar to those shown in Fig. \fr{exp}(b) could be directly observed. To enhance the contrast of the waves in such a demonstration setup a Fresnel lens could be used to collimate the light before it reflects off the rippled water surface. 

Furthermore, with an easy to implement synthetic Schlieren technique freely available online,\cite{Wildeman2018} precise measurements of the wave field can be acquired for digital post-processing (to obtain for example the intensity envelope). This Fourier transform based technique would also be fast enough for a real-time visualization.

Finally, the grating profile can be tuned with a sub-wavelength resolution in a straightforward manner. This ability could enable students to go a step further, and design, for example, gratings that endow the Talbot wave pattern with a fractal ingredient.\cite{Berry1996} Indeed, a nice introduction to another exciting field.

\begin{acknowledgments}
We thank Oded Agam, Steve Lipson and Antonin Eddi for fruitful and stimulating discussions. We thank Amaury Fourgeaud for his help in building the experimental setup. AB and SF would like to acknowledge partial support of Israel Science Foundation (ISF) grant 931/16. SW, EF and MF acknowledge the support of the AXA research fund and LABEX WIFI (ANR-10-LABX-24) within the French Program 'Investments for the Future' (ANR-10-IDEX-0001-02-PSL{*}).
\end{acknowledgments}

%\bibliographystyle{ajpdjg}
%\bibliography{talbotRefs}

\newpage

\begin{figure}
\centering
\includegraphics[width=\textwidth]{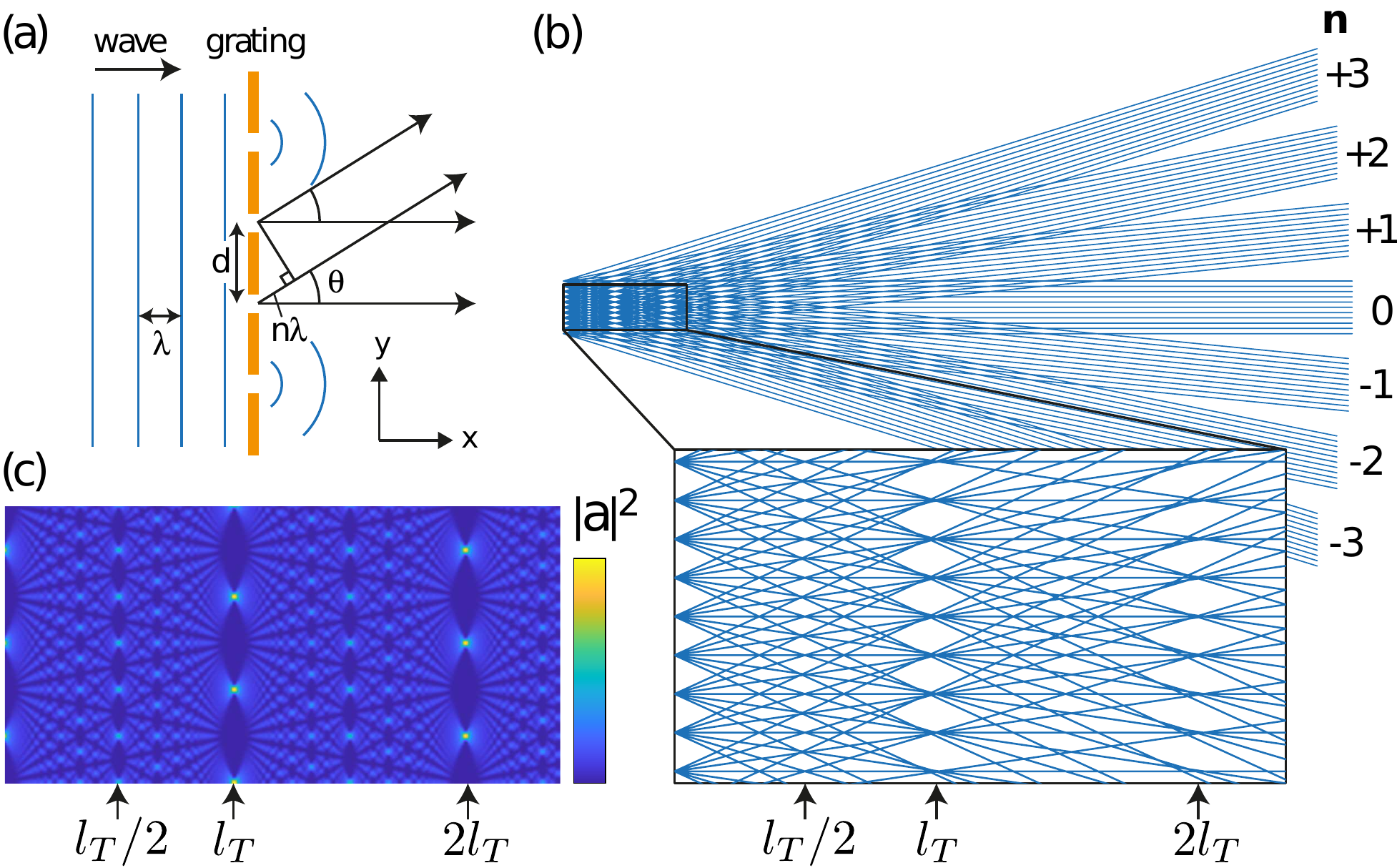}
\caption{Principle and model of the Talbot effect. (a) Basic setup. A monochromatic plane wave with wavelength $\lambda$ impinges orthogonally on a grating with period $d$. Constructive interference between waves originating from adjacent slits occurs if the path difference is $n\lambda$. (b) Geometrical interpretation of the Talbot effect, produced by the superposition of diffracting orders (drawn as rays from each slit). In the inset the self-imaging planes at once and twice the Talbot distance are clearly visible. Fractional Talbot images, at half the Talbot distance and with twice the spatial frequency, can also be distinguished. (c) Wave field intensity calculated using Eq. \er{series} for $\lambda/w = w/d = 0.1$.  Integer and Fractional Talbot images are visible.  In some of the focal planes interference effects cause the pattern to be shifted with respect to the geometric picture.}
\label{fig:theory}
\end{figure}

\begin{figure}
\centering
\includegraphics[width=\textwidth]{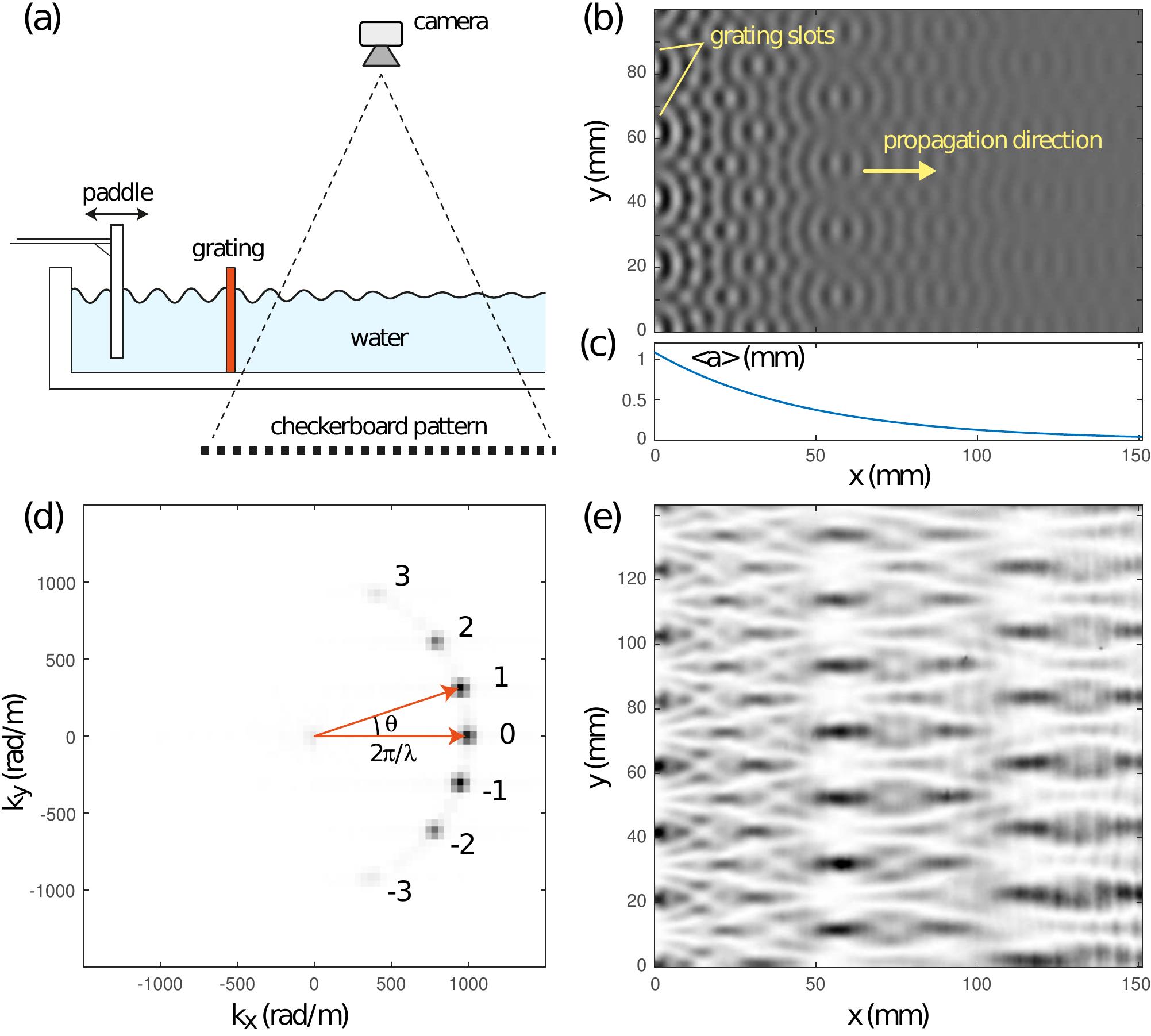}
\caption{Experimental method and data analysis. (a) Schematic of the setup. A paddle connected to a sub-woofer speaker (not shown) generates plane waves that pass through a periodic grating. A video camera, looking down through the water, records optical distortions in a checkerboard pattern placed below the basin. (b) Experimental wave field reconstructed from the checkerboard deformations. The grating slots and the wave propagation direction are indicated. For clarity of presentation the image is cropped around the five central slots. The same wave field repeats itself up to about two slots from the edge, where the finite size of the grating starts to play a role. (c) Average wave amplitude as a function of the distance from the grating. Due to viscous damping the wave amplitude decreases exponentially as it propagates away from the grating. (d) Spatial Fourier transform of the wave field. The diffraction orders show up as peaks laying on a circle with radius $|k| = 2\pi/\lambda$. (e) Intensity envelope of the wave field, corrected for the viscous decay.}
\label{fig:exp}
\end{figure}

\begin{figure}
\centering
\includegraphics[width=\textwidth]{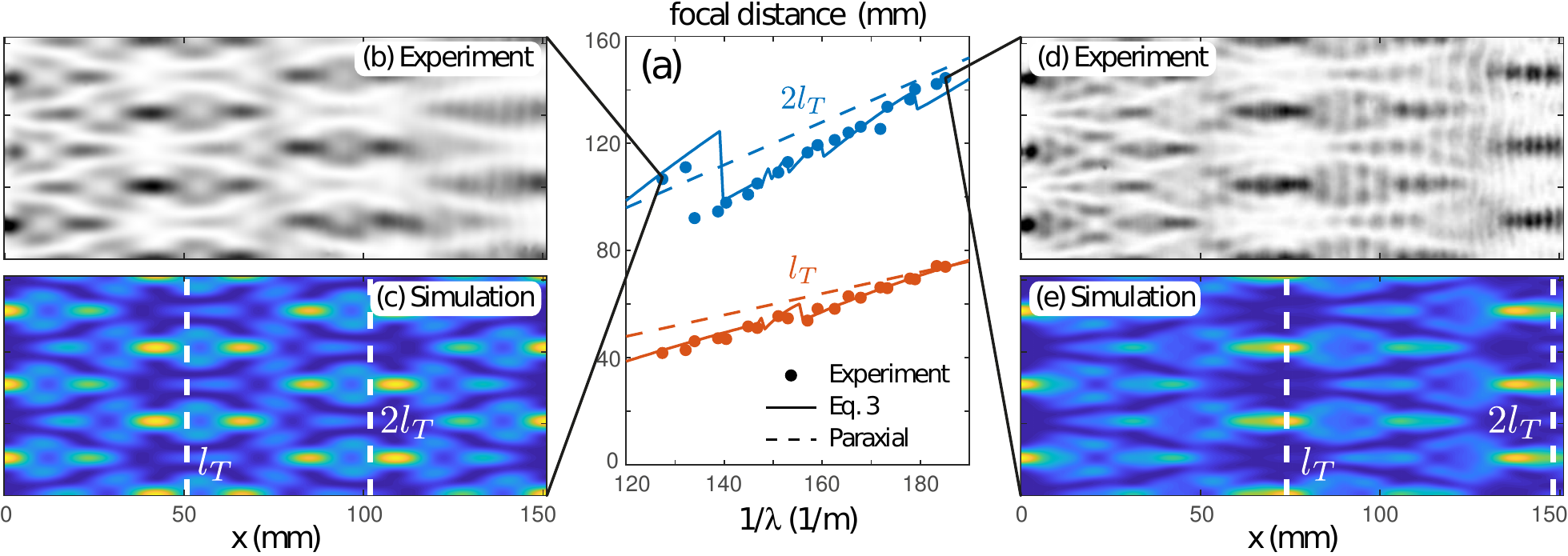}
\caption{Experimental results and comparison with the models. (a) Measured Talbot distances $l_T$ and $2l_T$ as a function of inverse wavelength, together with values obtained from Eq. \er{series} (solid lines) and the prediction of the paraxial theory (dashed lines).  Experimental distances are obtained by taking the distance to the grating where the variance of the traverse intensity profile reaches its local maximum. In (b) and (d) intensity profiles for the extreme experimental cases $\lambda = 7.8$\,mm ($f =  30$\,Hz) and $\lambda = 5.4$\,mm ($f =  49$\,Hz), respectively, are shown. Directly below that, in (c) and (e), the corresponding calculated profiles (using Eq. \er{series}) are shown for comparison. For clarity of presentation the vertically extended wave patterns are cropped around the three central slots. The vertical dashed lines indicate the theoretical Talbot distances according to the paraxial approximation.}
\label{fig:results}
\end{figure}

\end{document}